\documentclass[twocolumn,showpacs,preprintnumbers,amsmath,amssymb,epsf]{revtex4}

\usepackage{graphicx}
\usepackage{dcolumn}
\usepackage{bm}

\begin{document}


\title{Exit probability in a one-dimensional nonlinear q-voter model.}

\author{Piotr Przyby{\l}a, Katarzyna Sznajd--Weron and Maciej Tabiszewski}
\affiliation{Institute of Theoretical Physics, University of Wroc{\l}aw, pl. Maxa
Borna 9, 50-204 Wroc{\l}aw, Poland }

\date{\today}

\begin{abstract}
We formulate and investigate the nonlinear $q$-voter model (which as special cases includes the linear voter and the Sznajd model) on a one dimensional lattice.  We derive analytical formula for the exit probability and show that it agrees perfectly with Monte Carlo simulations. The puzzle, that we deal with here, may be contained in a simple question: ``Why the mean field approach gives the exact formula for the exit probability in the one-dimensional nonlinear $q$-voter model?''. To answer this question we test several hypothesis proposed recently for the Sznajd model, including the finite size effects, the influence of the range of interactions and the importance of the initial step of the evolution. On the one hand, our work is part of a trend of the current debate on the form of the exit probability in the one-dimensional Sznajd model but on the other hand, it concerns the much broader problem of nonlinear $q$-voter model.
\end{abstract}


\maketitle

\section{Introduction}
Linear voter model \cite{L1985}, one of the most recognized in a field of non-equilibrium phase transitions, is not only a toy model of an Ising spin system but also caricature of opinion dynamics. One of the main reasons of its importance is the fact that the linear voter model (VM) is solvable in arbitrary spacial dimension. However, from social point of view it is definitely to simplified and therefore several other models of opinion dynamics, based on Ising spins, have been introduced (for excellent resent review see \cite{CFL2009}), e.g. Sznajd model \cite{SWS2000} or majority model \cite{G1986,G2002}. Of course, it happens that a seemingly different models give the same results or even can be formulated in such a way that appear to be identical. For example, it has been shown that the original one-dimensional Sznajd model \cite{SWS2000} can be rewrite as a classical voter model \cite{BS2003}. However, the most commonly used version, in which only the unanimous pair changes the state of the system, differs significantly from VM. Nevertheless, it has been suggested that this case could be described by a broader class of nonlinear voter models \cite{SB2009,CMP2009}. 

Recently, particularly interesting nonlinear variant of the voter model, the $q$-voter model has been introduced \cite{CMP2009}. In the proposed model $q$, randomly picked (with possible repetitions), neighbors influence a voter to change opinion. If all $q$ neighbors agree, the voter takes their opinion; if they do not have a unanimous opinion, still a voter can flip with probability $\epsilon$. For $q=2$ and $\epsilon=0$ the model is almost identical with Sznajd model on a complete graph \cite{SL2003}. The only difference is that in the $q$-voter repetitions in choosing neighbors are possible. However, for $q=2$ and reasonable large lattice size this difference is negligible. In this paper we formulate and investigate the $q$-voter model on a one dimensional lattice for $\epsilon=0$. We show that analytical formula for the exit probability can be derived and the several approaches (among them the simple mean field approach) appear to lead to the same result. Moreover, received analytical formula agrees perfectly with Monte Carlo simulations. 
On the one hand, our work is part of a trend of the current debate on the form of the exit probability in the one-dimensional Sznajd model \cite{LR2008,SSP2008,CP2011,GM2011}. On the other hand, it concerns the much broader problem of nonlinear $q$-voter model. The puzzle, that we deal with here, may be contained in a simple question: ``Why the mean field approach gives the exact formula for the exit probability in the one-dimensional nonlinear $q$-voter model?'' 

We have to admit here that our question has been strongly inspired by the initial twofold question of Galam and  Martins "Why the mean field approach gives the exact formula for  the exit probability in your one-dimensional modified Sznajd model or why the Monte Carlo simulations give incorrectly a mean field result?". Their recent paper \cite{GM2011} is concluded with "Therefore the question is open for future work to settle either there is an explanation on why the system studied here exhibits a mean field behavior or why different simulations of the same system for different sizes all produced incorrectly a similar mean field result." We extend their question to generalized $q$-voter model and show by computer simulations which explanation of a puzzle given in \cite{GM2011} is more probable. 

\section{Nonlinear $q$-voter model in one dimension}
We consider a system of $L$ spins $S_i=\pm1$ located on a one dimensional ring. At each elementary time step $t$  a panel of $q$ neighboring spins $S_i,S_{i+1},\ldots,S_{i+q-1}$ is picked at random. If all $q$ neighbors are in the same state, they influence surrounding spins, if not all spins in the $q$-panel are equal then nothing changes. Two versions of the model are considered:
\begin{enumerate} 
\item
\textbf{Both sides:} The $q$-panel influences $R$ neighbors on the left side and the right side of the panel simultaneously -- all spins $S_{i-R}(t+\Delta t),S_{i-R+1}(t+\Delta t), \ldots S_{i-1}(t+\Delta t)$ and $S_{i+q+1}(t+\Delta t),S_{i+q+2}(t+\Delta t), \ldots S_{i+q+R-1} (t+\Delta t)$ take the state of the panel, i.e. $\rightarrow S_i(t)$. It is easy to notice that for $R=1$ and $q=2$ we deal with the original Sznajd model.  
\item
\textbf{Random:} The $q$-panel influences $R$ neighbors only on the one, randomly chosen, side (left or right) -- with probability $1/2$ spins $S_{i-R}(t+\Delta t),S_{i-R+1}(t+\Delta t), \ldots S_{i-1}(t+\Delta t) \rightarrow S_i(t)$ or ,with the same probability, $S_{i+q+1}(t+\Delta t),S_{i+q+2}(t+\Delta t), \ldots S_{i+q+R-1} (t+\Delta t) \rightarrow S_i(t)$. In this case for $R=1$ and $q=2$ we deal with the modified version of the Sznajd model, introduced by Slanina \cite{SSP2008}. For $R=1$ and $q=1$ we obtain original linear voter model \cite{KRB2010}.
\end{enumerate}
After one elementary step time increases by $\Delta t = 1/L$. Therefore time unit corresponds to one Monte Carlo Step (MCS). 

We have introduced both versions to be consistent with several other papers that deal with the problem of exit probability in the Sznajd model. In some of them ``both sides'' version is used \cite{CP2011}, whereas others deal with ``random'' version. However, as we will show, there is no difference between the exit probability for both versions and therefore each of them can be used.

\section{Exit Probability}
Let us consider a finite system with an initial fraction $\rho(0)$ of randomly distributed spins in the $+1$ state. Two scenarios for $q$-voter model on a one dimensional ring are possible:
\begin{itemize} 
\item
If there is no cluster of size $ \ge q$, the system is deadlocked and no evolution is possible, since only unanimous panel of size $q$ is able to change the state of the system. Obviously the number of deadlocks grows with $q$. For $q=1$ no configuration is deadlocked and for $q=2$ the only deadlocked configuration is antiferromagnetic state. 
\item
If there is at least one cluster of size $ \ge q$ then the system will evolve and eventually reach the ferromagnetic state -- with probability $E(\rho)$ the state 'all spins +1' is obtained, whereas with probability $1-E(\rho)$ the state 'all spins -1' is reached. $E(\rho)$ is called the exit probability and it is one of the most important first-passage properties \cite{KRB2010}.
\end{itemize} 

In 2008 the exit probability for the one-dimensional Sznajd model (which corresponds to nonlinear voter model with $q=2$) has been calculated analytically \cite{LR2008,SSP2008}:
\begin{equation}
E(\rho) = \frac{\rho^2}{\rho^2+(1-\rho)^2}.
\label{Ep2}
\end{equation}
This result agrees perfectly with Monte Carlo simulations, which is quite puzzling since calculations were not exact but based on  Kirkwood-approximation decoupling scheme \cite{LR2008,SSP2008}. Recently it has been shown that even much less sophisticated method, simple mean field approach, leads to the same result \cite{GM2011}. The question arises why the approximate methods give the exact result in this case? Some suggestions related to the importance of choosing the first pair appeared very recently \cite{CP2011,GM2011}.

In \cite{CP2011} the Sznajd model of range $R$ (SM(R)) has been studied. As usually, at each time step, a pair of nearest neighbors $S_i,S_{i+1}$ is chosen at random. If $S_i=S_{i+1}$ then $R$ neighbors to the left and $R$ neighbors to the right change value to $S_i$. This corresponds to the 'both sides' version of the nonlinear $q$-voter model (with $q=2$), introduced in the previous section. Remarkably, in the studied case the exit probability $E(\rho)$ turned out to be completely independent of the range of the interaction $R$. Based on this result $E(\rho)$ was derived in the following way -- the exit probability is given by the probability that a pair of sites in state $+1$ is chosen before any pair of sites in state $-1$ \cite{CP2011}:
\begin{eqnarray}
E(\rho) & = &  \rho^2 \sum_{n=0}^\infty \left[1 - \rho^2 - (1-\rho)^2 \right]^n \\
& = & \frac{\rho^2}{\rho^2+(1-\rho)^2}.
\end{eqnarray}

Another idea was presented in \cite{GM2011} -- the quasi-deterministic procedure in which the only random step
is the selection of the initial pair, and then the process is deterministic. These types of considerations have led again to the same formula for the exit probability (\ref{Ep2}).

In this paper we show that mean field approach gives the formula that agrees perfectly with Monte Carlo simulations for arbitrary value of $q$ and we also examine possible explanations why approximate methods give the exact result for a one-dimensional nonlinear $q$-voter model.

\begin{figure}
\begin{center}
\includegraphics[scale=0.4]{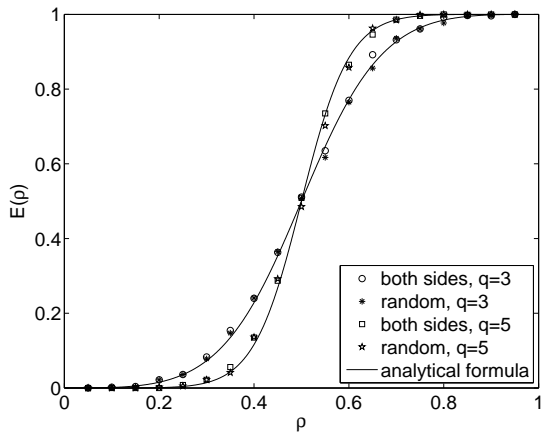}
\caption{Exit probability in the case of the ``random'' and ``both sides'' nonlinear q-voter model on a one-dimensional ring of length $L=100$. Averaging was done over $10^4$ samples. Results are identical for both versions of the model.}
\label{fig1}
\end{center}
\end{figure}

We start with showing by computer simulations that exit probability for both version of the model (``random'' and ``both sides'') are identical. In Fig. \ref{fig1} we present results for the lattice size $L=100$ and two values of $q$. Analogous results have been obtained earlier for the Sznajd model (i.e. $q=2$) \cite{SSP2008}. Given that both versions of the model give identical results, we will concentrate in the later work on ``random'' version. However, we have checked all results presented in this paper for both cases.

\begin{figure}
\begin{center}
\includegraphics[scale=0.4]{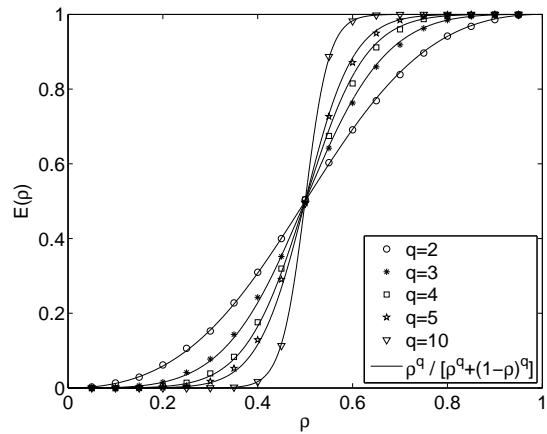}
\caption{Exit probability for nonlinear q-voter model on a one-dimensional ring of length $L=100$. Analytical formula agrees with the Monte Carlo results for any value of $q$. Averaging was done over $10^4$ samples.}
\label{fig2}
\end{center}
\end{figure}

Now we are ready to derive analytical formula for $q$-voter model. Let us denote by $<N_q^+(t)>$ the average value of the number of $q$-panels with all spins $+1$, and by by $<N_q^-(t)>$ the average value of the number of $q$-panels with all spins $+1$. Obviously:
\begin{eqnarray}
<N_q^+(t)>=\rho(t)^qL \nonumber \\ 
<N_q^-(t)>=(1-\rho(t))^qL,
\label{Nq}
\end{eqnarray}
where $\rho(t)$ is the fraction of spins $+1$ at time $t$.
Let us now introduce the quantity:
\begin{eqnarray}
m_q(t) & = & \frac{<N_q^+(t)>-<N_q^-(t)>}{<N_q^+(t)>+<N_q^-(t)>} \nonumber\\
& = & 2 \frac{\rho^q(t)}{\rho^q(t)+(1-\rho)^q(t)} -1,
\end{eqnarray}
which for $q=1$(linear voter model) is simply average magnetization:
\begin{eqnarray}
m_1(t) & = &\frac{<N_1^+(t)>-<N_1^-(t)>}{L} \nonumber\\
& = & 2\rho(t)-1.
\end{eqnarray}
It is known that for linear voter model average magnetization is constant, i.e. $m_1(t)=m_1(0)$ \cite{KRB2010}.
Knowing this, it is very easy to derive exit probability:
\begin{eqnarray}
m_1(\infty)=E(\rho) - (1-E(\rho))= 2E(\rho)-1,
\end{eqnarray}
where we use notation $\rho(0) \equiv \rho$.
Because
\begin{eqnarray}
m_1(\infty)=m_1(0)=2\rho-1,
\end{eqnarray}
we obtain that:
\begin{eqnarray}
E(\rho)=\rho.
\end{eqnarray}
Now we could generalize this reasoning assuming that:
\begin{eqnarray}
m_q(\infty) = m_q(0).
\label{ass_q}
\end{eqnarray}
With such an assumption we obtain result for arbitrary value of $q$:
\begin{eqnarray}
E(\rho)=\frac{\rho^q}{\rho^q+(1-\rho)^q}.
\label{Epq}
\end{eqnarray}
Indeed substituting to (\ref{Epq}) value of $q=1$ we obtain known result $E(\rho)=\rho$ and for $q=2$ we obtain formula
(\ref{Ep2}) derived independently in 4 papers \cite{LR2008,SSP2008,CP2011,GM2011}. Of course, since we did not show any proof that our assumption (\ref{ass_q}) is valid, the formula (\ref{Epq}) can be treated as a guess.
Let us start with checking validity of the formula (\ref{Epq}) by performing Monte Carlo simulations.
In Fig \ref{fig2} we present both, analytical and Monte Carlo, results for several values of $q$. As seen there is a perfect agreement though the number of averaging is not very large ($10^4$ samples).

\begin{figure}
\begin{center}
\includegraphics[scale=0.4]{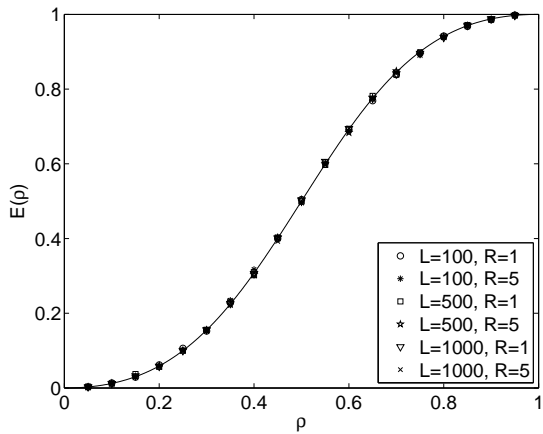}
\caption{Exit probability for nonlinear (q=2)-voter model on a one-dimensional ring of length $L$. Exit probability changes neither with the lattice size $L$ nor with the range of interactions. Analogous results has been obtained in \cite{LR2008,SSP2008,CP2011}. Analogous results are valid for any value of $q$. Averaging was done over $10^4$ samples.}
\label{fig3}
\end{center}
\end{figure}

Now we will follow the reasoning presented in \cite{CP2011}. We check how the exit probability depends on the range of interactions $R$. Up till now we focused only on $R=1$, but we have provided simulations also for $R=2,3,4,5$ and $10$. Results for $R=1$ and $R=5$ are presented in Fig. \ref{fig3}. Analogous results were obtained also for other values of $R$. It occurred that exit probability for the $q$-voter model is identical for any value of $R$. The same results has been obtained for the Sznajd model (i.e. $q=2$) in \cite{CP2011}. Remarkably, also the size of the system does not influence the exit probability (see Fig. \ref{fig3}). The same results have been obtained earlier for the Sznajd model \cite{SSP2008,CP2011}. This is quite intriguing, since in the case of a $q$-voter model on a complete graph the size of the system influences the exit probability  $E(\rho)$\cite{CMP2009}. Using the fact that the range of interactions does not change the exit probability we can derive analytical formula for $E(\rho)$ in the same way as in \cite{CP2011}.
For $R \ge (L-q)/2$ (in the case of ``both sides'') or $R \ge L-q$ (for ``random'') the system is fully ordered after first unanimous $q$-panel is chosen. Therefore the exit probability is equal the probability of choosing $q$-panel of 'up' spins:

\begin{equation}
E(\rho)=\frac{\rho^q}{\rho^q+(1-\rho)^q}.
\end{equation}

\begin{figure}
\begin{center}
\includegraphics[scale=0.4]{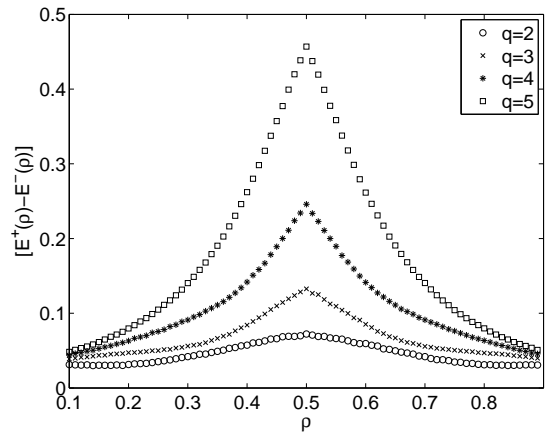}
\caption{Difference between two exit probabilities $E^+(\rho)$  and $E^-(\rho)$ for the lattice size $L=100$ and the range of interaction $R=1$. In the case of $E^+(\rho)$ a cluster of $q$ 'up' spins has been chosen in the first step of simulation, whereas in the case of $E^-(\rho)$ a cluster of $q$ 'down' spins has been chosen in the first step. It is seen that the first choice influence the exit probability. Averaging was done over $10^6$ samples.}
\label{fig4}
\end{center}
\end{figure}

\begin{figure}
\begin{center}
\includegraphics[scale=0.4]{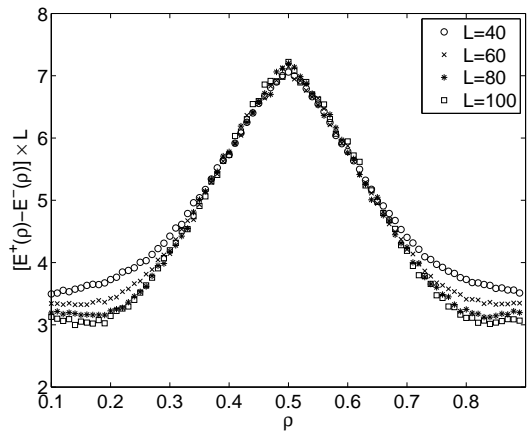}
\caption{Difference between two exit probabilities $E^+(\rho)$  and $E^-(\rho)$ for several lattice sizes $L$ and the range of interaction $R=1$. In the case of $E^+(\rho)$ a cluster of $q=2$ 'up' spins has been chosen in the first step of simulation, whereas in the case of $E^-(\rho)$ a cluster of $q=2$ 'down' spins has been chosen in the first step. Results scale trivially with the system size $L$. Averaging was done over $10^6$ samples.}
\label{fig5}
\end{center}
\end{figure}

This type of reasoning coincides, in a sense, the idea presented in the work of Galam and Martins \cite{GM2011}. They have proposed simple quasi-deterministic procedure that drives system to the absorbing ferromagnetic state with the same exit probability as in the Sznajd model. It their procedure the only probabilistic step is the choice of the first pair and further evolution is entirely deterministic. To see whether the first choice actually affects the probability of the final state we performed two types of simulations. In the first step we were picking at random $q$-panel of 'up' spins and then the system was evolving under standard procedure or in the first step we were picking at random $q$-panel of 'down' spins. We have denoted the exit probability in the first case by $E^+(\rho)$, whereas in the second case $E^-(\rho)$. If the first choice does not influence the exit probability we should obtain $E^+(\rho)=E^-(\rho)$. We have plotted in Fig. \ref{fig4} the difference $E^+(\rho)-E^-(\rho)$ for several values of $q$. As seen the difference grows with $q$, which is quite understandable. For large value of $q$ this is difficult to find unanimous $q$-panel, especially for $\rho \rightarrow 0.5$ (where only small clusters are present in the system). Probability of finding $q$-panel of 'up' spins is equal $\rho^q$, whereas probability of $q$-panel of 'down' spins is equal $(1-\rho)^q$. This explains  shapes of the curves in  Fig. \ref{fig4}. However, for $q=2$ importance of first choice is not too high and therefore quasi-deterministic procedure proposed by Galam and Martins \cite{GM2011} cannot explain the analytic formula for $E(\rho)$. Also the results concerning the importance of the interaction's range $R$ should be treated rather statistically. The fact that $R$ does not influence the exit probability does not mean that it is not important at all. In other case the first choice would determine the final state of the system, which is obviously not true (see Fig. \ref{fig4} and \ref{fig5}). 

\begin{figure}
\begin{center}
\includegraphics[scale=0.4]{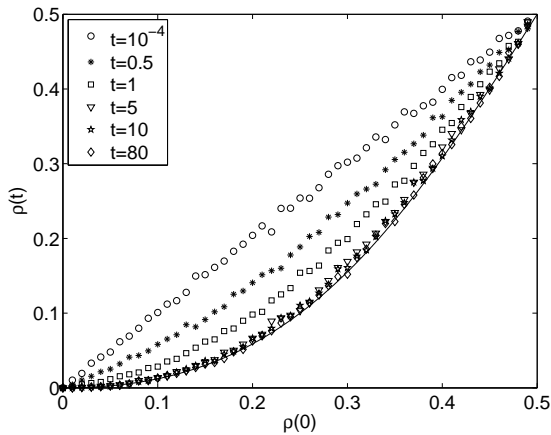}
\caption{Fraction of 'up' spins after time $t$ measured in the Monte Carlo Steps.}
\label{fig6}
\end{center}
\end{figure}

Although the choice of the first pair does not determine the result in 100\% but the first steps of simulation are actually the most important. To see this let us present probability $\rho(t)$ of choosing $q$-panel of 'up' spins as a function of initial probability $\rho$ for several values of $t$ (see Fig. \ref{fig6}). As we see, already after several Monte Carlo steps $\rho(t)$ concides with $E(\rho)$. Why the latter steps does not change the form of $E(\rho)$?

To understand this, we should mention here that the formula (\ref{Epq}) is valid only for random initial conditions. If we start from two clusters -- the first cluster of spins $+1$ and length $\rho L$ and the second one of spins $-1$ and length $(1-\rho)L$ we obtain the final state of all spins $+1$ with probability $\rho$. This is easy to understand because the probability in such a case of choosing $q$-panel of 'up' spins is equal $\rho$ (at least for the infinite system size $L$). The same result has been obtained for the Sznajd model in more sophisticated way using Kirkwood approach \cite{SSP2008}. Therefore, as soon as the system will order to several domains then $E(\rho)$ will no longer continue to change.

\section{Summary}
A recent work \cite{GM2011} did point out to the evoked puzzle emphasizing the mean field character of the formula for the exit probability in the Sznajd model obtained in \cite{LR2008,SSP2008}. It has been shown that the continuous shape of the exit probability is a direct outcome of a mean field treatment \cite{GM2011}. Two possible explanations has been given -- most likely finite size effects in the simulations or as a alternative irrelevance of fluctuations in the system.
In this paper we have extend the puzzle to the nonlinear $q$-voter model on a one-dimensional lattice. We have proposed analytical formula for the exit probability and checked it validity by Monte Carlo simulations for several values of $q$. In particular, we have shown that neither the range of interactions nor the size of the system influence the exit probability. Moreover, we have checked the importance of the initial evolution of the system, as suggested for the Sznajd model in \cite{GM2011,CP2011}, to understand proposed analytical formula. It should be stressed that results presented here contain a well-known cases of linear voter and Sznajd model.


\begin{thebibliography}{33}
\bibitem{L1985}
T.M. Liggett, \emph{Interacting particle systems} Springer, Heidelberg, 1985
\bibitem{CFL2009}
C. Castellano, S. Fortunato and V. Loreto , Rev. Mod. Phys. \textbf{81} (2009) 592
\bibitem{SWS2000}
K. Sznajd-Weron, J. Sznajd, Int. J. Mod. Phys. C \textbf{11} (2000) 1157
\bibitem{G1986}
S. Galam, Journal of of Mathematical Psychology \textbf{30} (1986) 426
\bibitem{G2002}
S. Galam, Eur. Phys. J. B \textbf{25} (2002) 403
\bibitem{BS2003}
L. Behera and F. Schweitzer,  International Journal of Modern Physics C \textbf{14}, 1331 (2003)
\bibitem{SB2009}
F. Schweitzer and L. Behera, Eur. Phys. J. B \textbf{67}, 301 (2009)
\bibitem{CMP2009}
C.Castellano, M.A.Muñoz, R.Pastor-Satorras, Phys. Rev. E \textbf{80}, 041129 (2009)
\bibitem{SL2003}
F. Slanina and H. Lavicka, Eur. Phys. J. B \textbf{35 }, 279 (2003)
\bibitem{LR2008}
R. Lambiotte and S. Redner, Europhys. Lett. \textbf{82}, 18007 (2008)
\bibitem{SSP2008}
F. Slanina, K. Sznajd-Weron, and P. Przyby{\l}a, Europhys. Lett. \textbf{82}, 18006 (2008)
\bibitem{CP2011} 
C. Castellano, R. Pastor-Satorras,  Phys. Rev. E \textbf{83}, 016113 (2011)
\bibitem{GM2011}
S. Galam and A.C.R. Martins, arXiv:1012.2283v1, to be published in Europhys. Lett. (2011)
\bibitem{KRB2010}
P.L. Krapivsky, S. Redner, E. Ben-Naim, \emph{A Kinetic View of Statistical Physics}, Cambridge University Press (2010)
\end{thebibliography}
\end{document}